Chapter

# Application of Process Mining and Sequence Clustering in Recognizing an Industrial Issue

*Hamza Saad*

**Abstract**

Process mining has become one of the best programs that can outline the event logs of production processes in visualized detail. We have addressed the important problem that easily occurs in the industrial process called Bottleneck. The analysis process was focused on extracting the bottlenecks in the production line to improve the flow of production. Given enough stored history logs, the field of process mining can provide a suitable answer to optimize production flow by mitigating bottlenecks in the production stream. Process mining diagnoses the productivity processes by mining event logs, this can help to expose the opportunities to optimize critical production processes. We found that there is a considerable bottleneck in the process because of the weaving activities. Through discussions with specialists, it was agreed that the main problem in the weaving processes, especially machines that were exhausted in overloading processes. The improvement in the system has been measured by teamwork; the cycle time for processes has improved to 91%. the worker's performance has improved from 96%, productivity has improved to 91%, product quality has improved by 85%, and lead time has optimized from days and weeks to hours.

## 1. Introduction

Process mining is a set of analysis techniques that explain the data-based overview of how a business process is executed in the real environment [1]. Typically, people who conduct business work have a obvious overview of how processes are working. Process mining performs the historical data to refute or confirm this belief. Process mining applications have confirmed in several cases, the output of conducting a process mining work can achieve real improvements in the process by depicting all activities and events in a dynamic map [2]. Process mining is a data-based approach to answer important question to reqonize business processes. That means the real data of the business process conduction must be plugged into an IT system, so a business activity includes many small task activities or subtasks activities. Thus, the IT system needs to record events that can be captured to these activities [3].

These activities require the case ID to reference the business process instances, the events are part of the activities, and timestamp represents the time when the event is executed. The events can hold relevant activities. Different types of process mining can be considered its main aims. Comprehensively, all types of process mining purpose to extract meaningful information in terms of many patterns from a set of process data logs being mapped and analyzed. According to Van der Aalst [4], process mining can be considered into the following three primary types: (i) business process conformance—which refers to comparing a business process model already known as an event log of the same business process to detect whether the reality, as recorded in the log, is in line with the business process model and vice versa; (ii) business process discovery—which refers to producing a still unknown business process model based on an event data log, using no prior information; and (iii) business process enhancement—which refers to modifying an existing business process model based on an event log of the same business process [5].

Process mining is entirely utilized in healthcare, business analysis, and service activities [6]. However, in the industry, process mining is still working to gain a good position as machine learning and data mining that have already been applied for decades. Process mining is established using algorithms of data mining and machine learning [7] so lastly, soft computing has been successfully applied, and the famous algorithm employed is a genetic algorithm [8]. There are many algorithms plugged into ProM software such as Fuzzy miner [9], Heuristic miner [10], and the Genetic miner [11]. However, some algorithms are unable to provide high performance to confirm if there is a significant bottleneck in the process, build tracks among activities, or improve the process. Some algorithms of soft computing can predict a proper sequence for activities in the process, calculate exact frequency, and calculate the exact idle time among activities in the same process mining diagram. Data in the industrial field includes many processes that need to be more focused on, and this cannot be done by studying the whole system but can be done by clustering the data into a number of homogeneous splits.

If all the required data is uploaded into the process mining algorithms, then the data will be extracted, and preprocessed and the event logs are constructed. An event data log is a group of events that follow the same business process. It is the implied data type upon which all process mining algorithms are constructed. Once the event data log is conducted, the process mining technique can be applied to recognize data.

As with many studies, process mining is relatively still growing. To limit it, one can say it is limited by soft computing, machine learning, and data mining on one side and by analysis and process modeling on the other side [4]. It has acquired some conductions by different industries over the last years. Some examples of software for process mining are Celonis PI and Fluxicon Disco. Disco is a more straightforward software, and it can accept big data based on the license that is provided by the company. Using this software does not require much knowledge to interpret results. There is no previous use of data mining in the textile field, especially for this case study. ProM has many algorithms plugged in, but not all algorithms are doing well with this data.

The study explains the evaluation method based on process mining to overcome bottlenecks in the production processes. The method is based on a defined framework, guidelines, and methodologies for process mining projects [12, 13], and adopted a question-driven process mining project (the management will be asked to confirm and suggest the changes in the process). It provides the main stages of a production process evaluation to improve the understandability and usability of process mining in the unstructured process for non-experts [14]. The proposed process mining methodology searches the ways of displaying simple processes by splitting data into smaller similar datasets and focusing on the specific similarity clusters.

| Action | Statistics |
| --- | --- |
| Events | 443 |
| Cases | 33 |
| Activities | 16 |
| Median case duration | 43.4 Weeks |
| Mean case duration | 49.7 Weeks |
| Start | 01/01/2019 01:00:00 |
| End | 05//06/2021 19:12:00 |

**Table 1.**
*Statistics from Event-Logs for whole process.*

## 2. Data exploration using process mining

Data is collected regarding cases, activities, units, shifts, and timeframe. In this process, the event data is automatically extracted from the processes of the company job shop. Because the process mining had not been applied before in textile production, especially in the country of the case study, we spent 6 months waiting to get this data in the form of event logs. It contains information about 443 events, 33 cases, and 14 activities in the real production process. The data is stored as CSV Excel sheets and flat files. The real-world event log is formatted as MXML (Mining Extensible Markup Language) using the ProM tools [15], and (CVS Excel sheet) when Disco process mining is applied to solve data. **Table 1** shows the statistics of event logs after process mining is applied to solve data.

The method is proposed to control, monitor, and motivate the improvement of textile production processes and is based on the principles of the goal-driven process mining project [12]. The ultimate goal of the study is to investigate the applicability of process mining and its potential in improving the process of textile production.

## 3. Methodology

Agile project management has been followed to build the methodology to improve the process in production. The short-limit goal was designed to improve the current production process by removing any unnecessary work and extracting the important knowledge to manage the process in the job shop. The problem has been defined, and based on this, data in the form of event logs has been collected to fit process mining applications. Data was collected from the company archive and current job shop processes. The collected data was heterogeneous with many missed and high dimensional data. It has been filtered and preprocessed for use in Disco and ProM. Disco software has professional visualization and analysis of data, but this program has only one option to analyze data. Many algorithms from data mining, machine learning, soft computing, and fuzzy logic are plugged into the Platform of ProM. In addition, this platform has open access to plug in any algorithm's updates in process mining. The data form has been converted to be accepted in ProM. Sequence clustering based on the Markov Chain in ProM was used to split the original data into three splits. So, ProM has basic visualization, thus clustered data is returned to Disco to get more details about the bottlenecks in the process by analyzing each cluster. By getting the results from clustered data and comparing it with the original data, the initial evaluation has been built for the management. The main issue was in the weaving processes with massive bottlenecks. Using agile project management and lean principles, some features have been chosen to change management and improve the process. Some lean principles that are more fitted to apply in the job shop have been estimated, like cellular manufacturing and VSM in the production line, and leadership training for the supervisors in the production line. We can apply VSM to visualize the process, but we found some limits to VSM, and these limits have been overcome by process mining when visualizing the whole process and extracting the main bottlenecks. We worked hard to close the project on time, so after 5 months we got improvement in the process. The methodology development is presented in **Figure 1**.

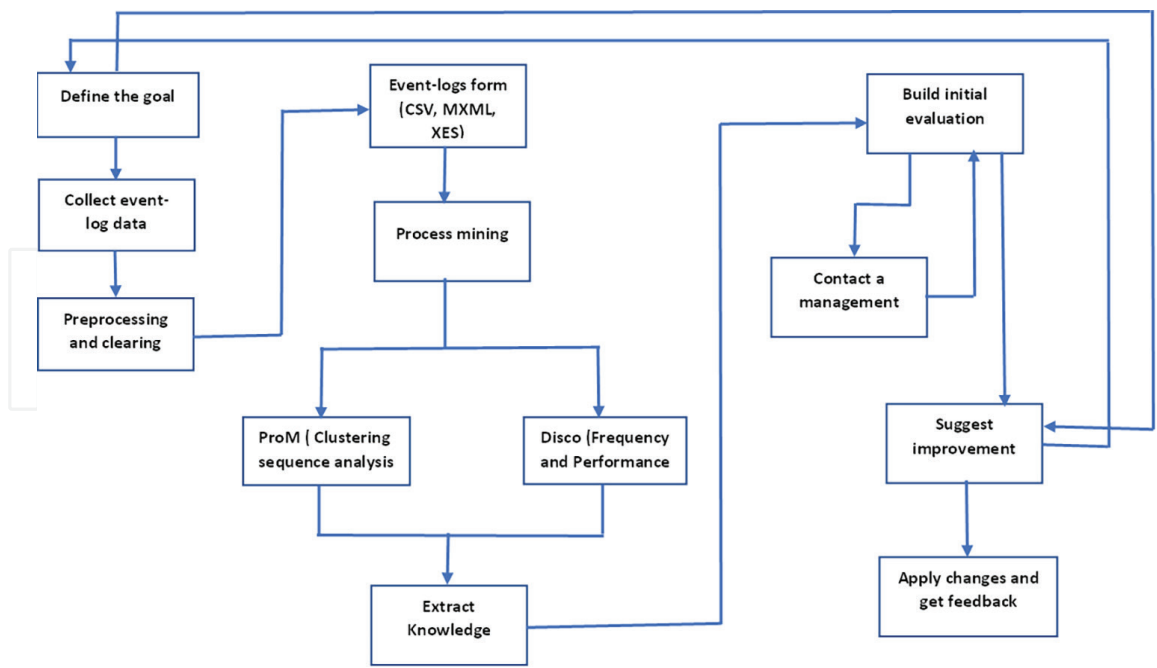

**Figure 1.**
*Methodology development.*

## 4. Data preprocessing and analyzing

There are many algorithms of data mining applied to cluster the data such as K-means and Mean-Shift Cluster. However, in the process mining, some activities follow one track on the map, this agreement is called a variant. It is not entirely like the cluster technique in data mining, but it can be used to solve the problem based on homogeneous activities. Data has 25 variants based on the number of activity paths in the whole map. The original period that was picked to collect this data was 5 months starting from 01/01/2019 to 05/31/2019, but the algorithms of process mining to provide an understood and appropriate map generated the process for two years and 93 days which begins from 01/01/2019 and finished on 06/05/2021. Variant 2, which has 11 activities recorded the lowest time duration (5 hours, 55 minutes), whereas the highest time duration recorded by variants 7 and 10 which have respectively 14 and 16 activities, both have the same time duration of 2 years and 93 days.

Process mining uses event logs to explain the real behavior in the conduction of the business process. Event logs consist of a set of traces that represent the process instance that began in the information system. Each trace contained one activity track, so it is called a process case. A case is typically traceable by one or more fields known as a case identifier. The process mining identified which activity has a high frequency and long idle performance.

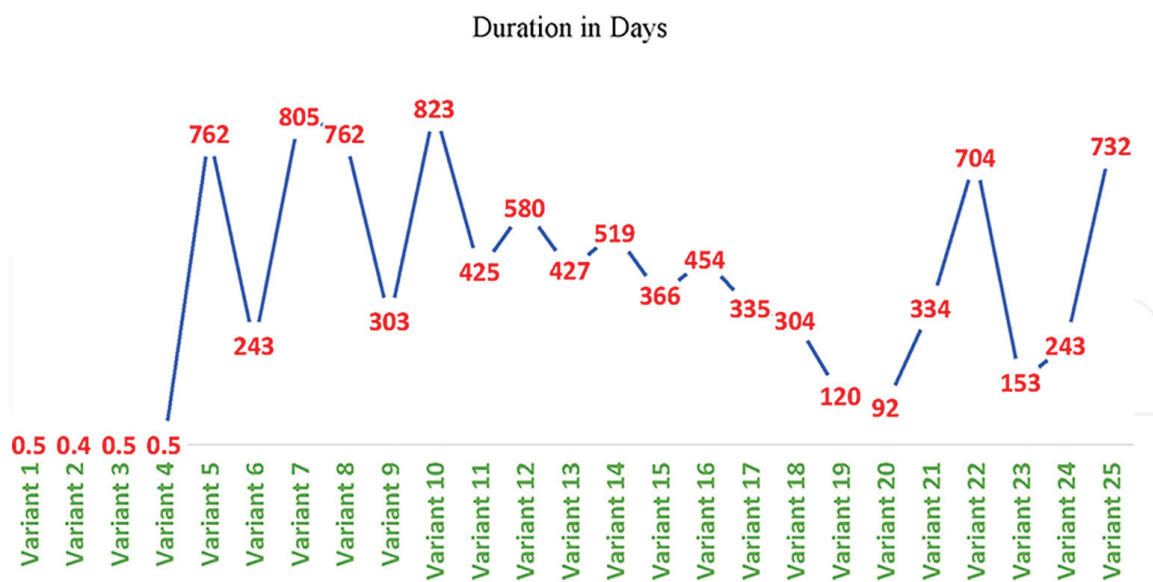

**Figure 2.**
*Time duration for each variant.*

Each track or path includes some activities, and each activity takes a specific time to be completed before starting another one. **Figure 2**. Variants and time duration.

The results of the process mining using Disco are explained in 25 variants; each variant represents one full path or track for groups of events or activities on the map (the activitities iclude part of production steps). Each variant has random activities and a specific time duration.

The majority of frequency goes to weaving activity which has the main bottleneck in the whole map of process mining based on the management answers. Process mining cannot provide a complete solution to the management, but it can provide explicit detail about what and where is the problem in the real process. Workers in the real system and data analyzers need to work together to remove bottlenecks and improve the work process. **Table 2** shows the frequency of each activity.

Frequency shows the most recurred process which is located at the weaving process with 162 frequencies. The only important absolute frequency is located at the weaving activity. However, in the case of frequency, there are many important activities like weaving, final shape, and drawing.

Furthermore, the bottleneck is entirely exposed in the tracks of weaving and weaving activity with the total duration reaching 66.1 months.

On the other hand, the bottlenecks exposed in most tracks with the huge time durations. The initial decision is that there is an abnormal activity in the weaving process and the question that should be answered by management is what the work efficiency and machine productivity in the textile department are. Moreover, how does the textile process relate to the rest of the processes in the whole process of production? Performance mode in Disco measures the time that is not utilized in the process (Idle time) because the process waits a while until the next activity starts its job. In the total duration activity, the total duration between the same activities in the textile process is 66.1 months. It is a significant bottleneck exposed in one weaving activity from total duration performance. However, by changing the mode from total duration to mean duration, the bottlenecks were exposed in three locations; a small bottleneck between sample testing and washing with a total duration of 39.7 weeks, a bottleneck between washing and silver package with a total duration of 51.4 weeks, and bottleneck between blending and reeling with total duration is 43.7 weeks.

| Activity | Frequency | Relative frequency (%) |
| --- | --- | --- |
| Weaving | 162 | 36.57 |
| Sample testing | 41 | 9.26 |
| Drawing | 25 | 5.64 |
| Final shape | 25 | 5.64 |
| Silver package | 24 | 5.42 |
| Winding stage | 23 | 5.19 |
| Fine spinning | 23 | 5.19 |
| Twisting | 23 | 5.19 |
| Assembly winding | 23 | 5.19 |
| Reeling | 23 | 5.19 |
| Dying | 17 | 3.84 |
| Washing | 13 | 2.93 |
| Blending | 13 | 2.93 |
| Raw wool receiving | 8 | 1.81 |

**Table 2.**
*Activity and its frequency in the process (these are all activities in the process ranked based on the highest frequency).*

There is a difference between total duration performance and mean duration performance. In the mean duration, there is no bottleneck exposed on weaving activity because data of event logs was clarified based on the clusters to compare the number of clusters instead of performance duration. Virtually, there was no doubt, because all results and bottlenecks that were exposed in the process mapping were sent to the management to give final suggestions and give details about the process to solve the problem, improve productivity, and estimate the next step to improve the process.

## 5. Sequence clustering

Sequence clustering is a machine learning technique that takes several sequences and gathers them in clusters so that each cluster group in similar sequences. The development of these methods has active research and studies, especially in connection with challenges in the field of informatics [16] and healthcare [17]. A simple sequence clustering algorithm is based on first-order Markov chains [18]. In the algorithm, every group cluster is established with the first-order Markov chain, where the current state depends only on the previous state. The probability that observed sequences belong to a presented group cluster is in effect the probability that the observed sequence was generated by the Markov chain assigned with that cluster [19].

For a sequence x = { $x_0$, $x_1$, $x_2$,..., $x_{L-1}$ } of length L this can be simply explained as:

$$P\left(\frac{x}{c_k}\right) = P(x_0, c_k) \cdot \prod_{i=1}^{i=L-1} P\left(\frac{x_i}{x_{i-1}}, c_k\right) \quad (1)$$

$P(x_0, c_k)$ refers to the probability of $x_0$ occurs at the first state in Markov chain assigned with the cluster $c_k$. $P\left(\frac{x_i}{x_{i-1}}, c_k\right)$ refers to the transition state probability $x_{i-1}$ to the state $x_i$ in the same Markov chain. By giving the information to calculate $P\left(\frac{x}{c_k}\right)$, the algorithm of sequence clustering can be conducted as an extent to the well-known Algorithm of Expectation-Maximization [20]. The steps for the Markov chain cluster are:

1. Initialize randomly the model parameters $P\left(\frac{x_i}{x_{i-1}}, c_k\right)$ and $P(x_0, c_k)$. For every group cluster, the probabilities of state transition of the assigned Markov chain are randomly initialized.

2. Use the parameters of the current model, define each sequence to each cluster with probability 1.

3. To re-estimate and calculate the model parameters in the next step, use results in step 2 to recalculate the probabilities of state transition for every Markov chain according to the sequences that agree to that cluster.

4. Repeat steps 2 and step 3 until the mixture model gets together.

The cluster is conducted in a solution based on three splits in the original dataset, 0, 1, and 2. Cluster 0 includes 164 events, cluster 1 has 117 events, and cluster 2 includes 229 events, the summation for these events is 510. However, 510 is not equal to the original total events that equal to 443 because these clusters or some clusters do not include pure split due to some events recurred in other clusters. The main map is generated from Disco process mining and Markov Chain in ProM process mining, the clusters from the Markov chain are:

Cluster 0 includes 11 instances (**Table 3** and **Figure 3**).
Cluster 1 includes 5 instances (**Table 4** and **Figure 4**).
Cluster 2 includes 15 instances (**Table 5**, **Figure 5**).

| Instances | Cases | Events | Duration |
|---|---|---|---|
| 3 | 1, 11, 32 | 18 | 60 days |
| 2 | 2, 23 | 13 | 28 days |
| 2 | 3, 24 | 11 | 13 days |
| 1 | 10 | 18 | 2 days |
| 1 | 14 | 16 | 1 day |
| 1 | 15 | 11 | 10 h |
| 1 | 31 | 16 | 5.23 h |

**Table 3.**
*Instances, events, and cases in cluster 0.*

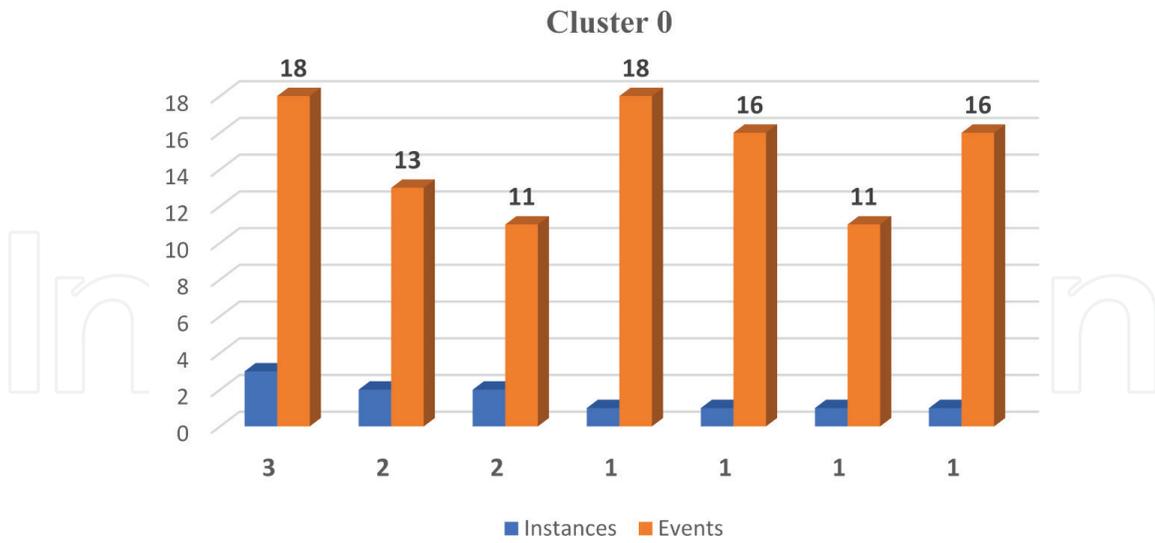

**Figure 3.**
*Cluster 0 results.*

| Instances | Cases # | Events | Duration |
|---|---|---|---|
| 2 | 4, 25 | 17 | 45 days |
| 2 | 6, 17 | 16 | 32 days |
| 1 | 8 | 17 | 17 days |
| 1 | 12 | 13 | 8 days |
| 1 | 33 | 21 | 2 days |

**Table 4.**
*Instances, events, and cases in cluster 1.*

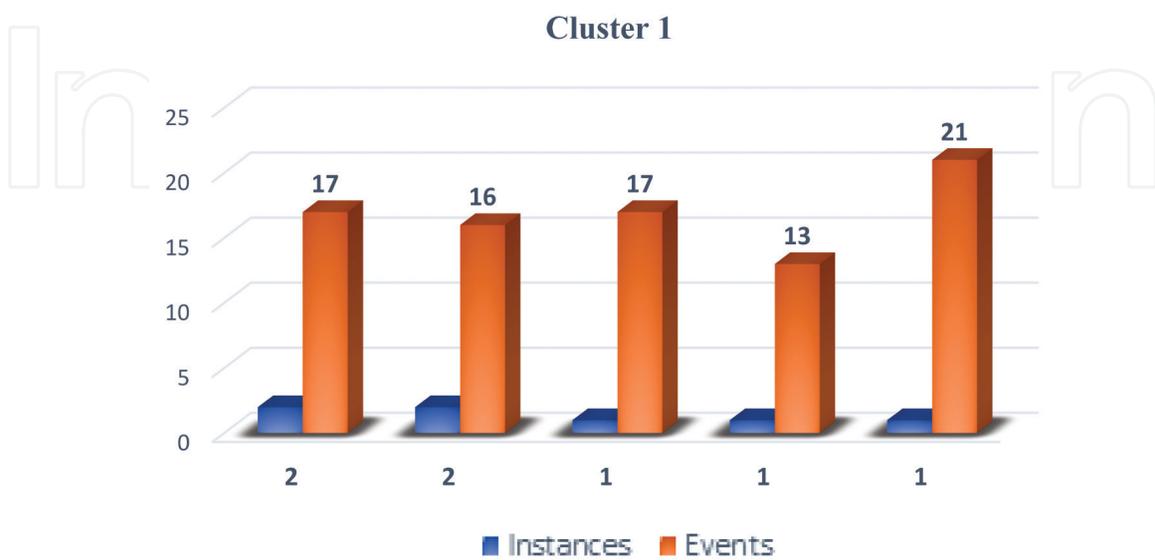

**Figure 4.**
*Cluster 1 results.*

| Instances | Cases | Events | Duration |
| --- | --- | --- | --- |
| 2 | 5, 26 | 18 | 64 days |
| 2 | 7, 28 | 11 | 51 days |
| 1 | 9 | 25 | 32 days |
| 1 | 13 | 11 | 29 days |
| 1 | 16 | 22 | 23 days |
| 1 | 17 | 16 | 22 days |
| 1 | 18 | 13 | 22 days |
| 1 | 19 | 15 | 17 days |
| 1 | 20 | 14 | 22 days |
| 1 | 21 | 16 | 11 days |
| 1 | 22 | 16 | 11 days |
| 1 | 29 | 9 | 5 days |
| 1 | 30 | 12 | 7 days |

**Table 5.**
*Instances, events, and cases in cluster 2.*

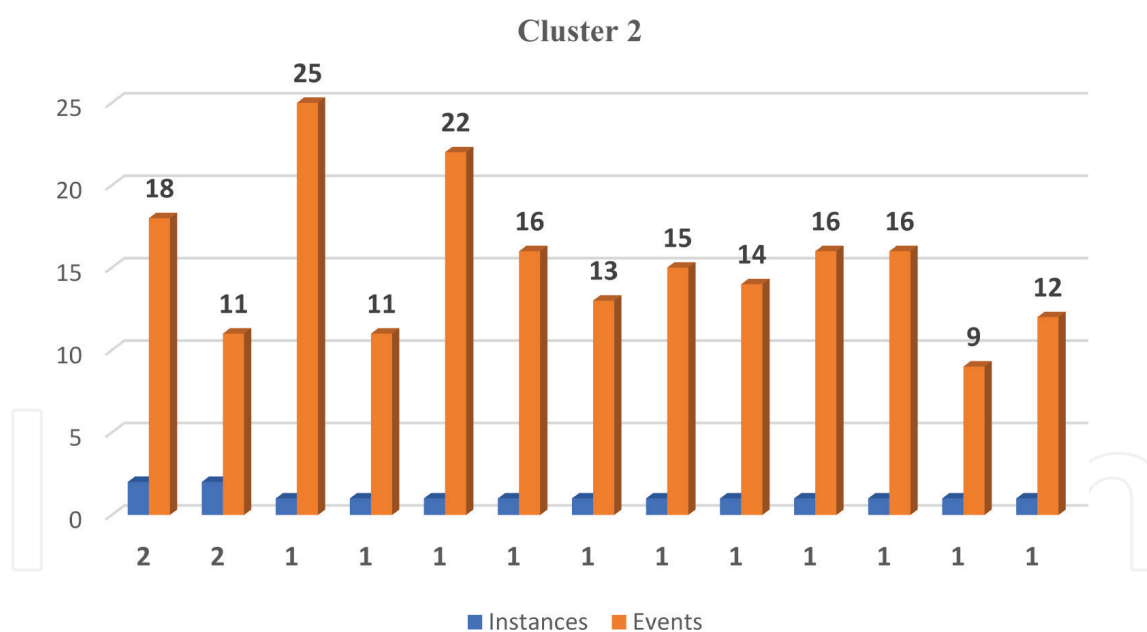

**Figure 5.**
*Cluster 2 results.*

## 6. Discussion

Clustering using Markov chain in ProM is not clear enough to find bottlenecks or exact idle times, but to solve this problem, clustered data has moved to another software called Disco by respecting each cluster to extract more information can help to improve the current processes, so Disco software can give full details about each cluster by mapping the activities and giving full process details.

At cluster 0, the huge idle time is from assembly wending to weaving. The total duration is 61.3 days. The high relative frequency recorded to weaving activity is 22.09%.

At cluster 1, the huge idle time is from silver package to weaving. The total time duration is 52.1 weeks. The high relative frequency recorded to weaving activity is 28.21%.

At cluster 2, the huge idle time is from weaving to weaving. The total time duration is 38 months, so the high relative frequency recorded to weaving activity is 40.61%. The time from blending to sampling testing is huge also, but management said that time is reasonable because the tests are not always required if productivity is in the same design.

The cluster is playing a vital role in extracting more information by splitting data into homogeneous groups. Also, Disco does not have a clustering technique, but it can show the impact and the bottleneck by using the variants. These variants are close to clusters with more details because they give details for each track or bath in one variant that includes many activities, based on how much data that uploaded into the system.

By taking clustered data that was clustered using sequence clustering in ProM and uploading it to the Disco program to provide more details about data; cluster 0 has no bottleneck based on the performance of the total duration. Cluster 1 includes a long idle time reaching 52.1 weeks between the silver package and weaving activity; besides that, other occasional bottlenecks do not affect the process. However, in cluster 2 the algorithm failed to match all activities in the one integrated path because data in this case included a high dimensional clustered dataset, so in the split process, there is a bottleneck between blending and sample testing and the idle time between those activities is 51.2 weeks, but the considerable bottleneck is between Weaving and Weaving activity which its idle time is 38 months for same activity.

## 7. Process improvement

We applied the changes and waited time to evaluate the changes in the process as presented in **Table 6**.

The scope of the project focused on the main bottlenecks in the weaving processes. The deliverables of the project were optimizing the main bottlenecks by considering a group of factors to improve the production process. These factors are improving worker's efficiency, reducing lead time, reducing cycle time, and quality improvement. All training is planned to be done in 1 month to start getting improvement in the process. The people who work in the job shop have managed to work on some tasks according to agile project management until completing the final project. In the brainstorming meeting, the team scheduled leadership training for worker efficiency, reducing lead time and cycle time by applying VSM and Kaizen events, productivity, and machines management by cellular manufacturing and visualization management.

Agile project management helped a lot to finish the training in 1 month by prioritizing the most important features to finish on time. The changes in the system have been measured by the team to confirm these improvements; the worker's performance improved from 59 to 96% after 3 months, lead time has optimized from days and weeks to days and hours, productivity has improved from 65 to 91%, product quality has improved 80% by building quality assurance system, and cycle time for all process has improved to 91%.

*Application of Process Mining and Sequence Clustering in Recognizing an Industrial Issue*
*DOI: http://dx.doi.org/10.5772/intechopen.113843*

| Item | Status | Method applied | process of improving | Result after 5 months |
|---|---|---|---|---|
| Bottlenecks | Big bottleneck in weaving unit | VSM | Redistribute machines and add three more in production line | Bottleneck has optimized by considering more other factors |
| Worker efficiency | Unskilled and irresponsible | Lean principles and leadership training | Training, and hire professional supervisors to manage the process | Efficiency has improved with 37% after 3 months |
| Process lead time | Main and occasional bottlenecks | Kaizen events, and VSM | Improve the efficiency of workers and production line | Time is optimized from 3 h. to 1.23 h. After 3 months |
| Productivity | Machines, and workers | Lean, cellular manufacturing, and visualization management | Redistribute machines in the production line and hire professional workers | The productivity improved from 65% to 91% after 5 months |
| Production process | Stops of machines, and some machines are old and slow | Kaizen events and VSM | Apply a complete change in the process and create updated work standardization. | Production became faster |
| Quality | Irregularity of work | Quality assurance | Remove sources of defects | Improved to 85% |
| Lead time and cycle time | Long lead and cycle time | Swim lane designed to eliminate unnecessary work | Work schedule and improve productivity | Lead time has reduced from months to days, and cycle time relates to the productivity that improved 91% |

**Table 6.**
*Apply change management to improve the process.*

## 8. Conclusion

Process mining is applied to improve industrial productivity by optimizing bottlenecks. Disco is used to analyze the original data to find a big bottleneck in the process. So, we decided to extract more information by splitting the original dataset. Disco has only one option to analyze data, but there are many options can we get from ProM. Each software has specific requirements to accept data. The form of data has been converted to fit ProM. Sequence clustering-based Markov chain has been used to split data. Because ProM has bad visualization, the clustered data has moved to Disco to visualize and analyze data based on each cluster. By analyzing clusters, huge bottlenecks have been found in the weaving processes. The agile project management is considered to tackle these bottlenecks by considering some factors that can play a vital role in improving the process. So, teamwork got internal training for lean principles to manage project tasks. After 5 months, The improvement in the system have been measured by teamwork; the cycle time for all process has improved to 91%. the worker's performance has improved from 96%, productivity has improved to 91%, product quality has improved by 85%, and lead time has optimized from days and weeks to hours.

## Author details

Hamza Saad
Technology, Art and Design Department, Bemidji State University, MN, USA

*Address all correspondence to: hamza.saad@bemidjistate.edu

IntechOpen